\documentclass[%
twocolumn,nofootinbib,amsmath,prd,aps,superscriptaddress,
tightenlines,preprintnumbers,nobalancelastpage]{revtex4-2}

\usepackage{graphicx}
\usepackage{dcolumn}
\usepackage{bm}
\usepackage[utf8]{inputenc}
\usepackage{ragged2e}
\usepackage{calligra,amsmath,amsfonts,bbm,mathrsfs,amssymb,mathtools,amsthm}
\usepackage{slashed,cancel}
\usepackage{hyperref}
\hypersetup{colorlinks=true,urlcolor=blue,anchorcolor=blue,citecolor=blue,filecolor=blue,
            linkcolor=blue,menucolor=blue, linktocpage=true,pdfproducer=medialab}
\usepackage[english]{babel}
\usepackage{indentfirst}
\usepackage{float}
\usepackage{enumerate}
\usepackage{subcaption}
\usepackage{color}
\usepackage{multirow}
\usepackage[normalem]{ulem} 
\usepackage{booktabs}
\usepackage{braket}
\usepackage[dvipsnames]{xcolor}
\usepackage{MnSymbol}
\usepackage[skins,theorems]{tcolorbox}
\usepackage[capitalise]{cleveref}
\usepackage{orcidlink}
\usepackage[font=small,labelfont=bf,justification=centerlast]{caption}
\tcbset{highlight math style={enhanced,
  colframe=red,colback=white,arc=0pt,boxrule=1pt}}
\newcommand{\lgr}{\mathcal{L}}

\newcommand{\Mpl}{M_{\rm Pl}}

\renewcommand{\d}{\text{d}}

\numberwithin{equation}{section}

\makeatletter

\newcommand{\edit}[1]{{\color{black}{#1}}}

\begin{document}

\title{\Large Misalignment from kicks: the impact of particle interactions on ultra-light dark matter}

\author{Clare Burrage}
 \email{clare.burrage@nottingham.ac.uk}
\affiliation{School of Physics and Astronomy, University of Nottingham, University Park, Nottingham, NG7 2RD, UK}

\author{Sergio Sevillano Muñoz}
\email{sergiosm@sas.upenn.edu}
\affiliation{Institute for Particle Physics Phenomenology, Durham University, South Road, DH1 3LE, Durham, UK}
\affiliation{Center for Particle Cosmology, Department of Physics and Astronomy, University of Pennsylvania, Philadelphia, Pennsylvania 19104, USA}

\begin{abstract}
Oscillating ultra-light scalar fields are a natural explanation for the dark matter in our universe, as long as a mechanism, often called a misalignment mechanism, exists to explain the amplitude of the scalar oscillations. If the dark matter scalar couples to the Standard Model, then the dynamics of ordinary matter can influence the behaviour of dark matter in the early universe. In this work we show how this changes the expected value of the scalar field and the resulting amplitude of late time scalar oscillations, and therefore the abundance of dark matter at late times. For dark matter scalars that interact quadratically with Standard Model fields we derive estimates of the size of this effect as a function of the strength of the coupling, and for axion-like fields we show that interactions with dark sector matter can temporarily destabilize the field, leading to large field displacements. 
\end{abstract}

\maketitle

\section{Introduction}

Ultra-light scalar fields are a popular explanation for the fundamental nature of the dark matter that makes up 27\% of our Universe today \cite{Planck:2018vyg}. If such a scalar field is oscillating homogeneously in a quadratic potential, then it behaves as a pressure-less perfect fluid \cite{Preskill:1982cy,Abbott:1982af,Dine:1982ah,Hu:2000ke, Hui:2021tkt, Antypas:2022asj}. As long as the scalar is not too light, $m_\phi \gtrsim 10^{-24} \mbox{ eV}$ \cite{Ferreira:2020fam}, then the scalar is also able to cluster and form halos around visible galaxies. 

One challenge for ultra-light dark matter models is to explain the observed amount of dark matter. The dark matter abundance in these models is set by the mass of the scalar, which controls the frequency of oscillations, and the amplitude of oscillations. So an ultra-light scalar explanation for dark matter requires an answer to the question: Why is the scalar field oscillating in its potential, and not settled at the minimum?  Solutions to this problem are referred to as the {\it misalignment mechanism} \cite{Ferreira:2020fam,OHare:2024nmr}, because the value of the scalar field must, at some time in the early universe, be misaligned from the position of the minimum of its potential at late times. The scalar field will be frozen by Hubble friction, at this misaligned value until a time at which the Hubble scale drops below the mass of the scalar, after this point the field will oscillate in its potential. 

Explanations for the initial misaligned value of the scalar often invoke an additional symmetry which is broken to give rise to the misaligned initial value of the scalar. For example the scalar could be the pseudo-Nambu Goldstone boson that emerges when a $U(1)$ symmetry is spontaneously broken at a high energy scale and then explicitly broken, e.g. by temperature dependent corrections as occurs for axion models \cite{Batell:2021ofv, Batell:2022qvr, OHare:2024nmr, Cyncynates:2024ufu, Cyncynates:2024bxw, Ferrante:2025avp, GuptaChoudhury:2025uje}.  In this work, we show how misalignment can be enhanced or suppressed by interactions with Standard Model fields. 

We focus on dark matter scalar fields with quadratic couplings to the Standard Model.  There has been recent interest in such models
in part because they avoid many of the stringent constraints on linear couplings \cite{Hees:2018fpg,Lee:2020zjt}, and in part because such interactions occur for axions \cite{Bauer:2023czj,Grossman:2025cov}, symmetric Higgs portal scalars \cite{OConnell:2006rsp,Patt:2006fw}, or other scalar models where linear terms are forbidden by some mirror symmetry \cite{Delaunay:2025pho}. Such scalars are constrained by measurements with atomic clocks, and by the MICROSCOPE  satellite's test of the equivalence principle \cite{Kim:2022ype,Panda:2023nir,Hees:2018fpg,Filzinger:2023qqh,Elder:2025tue,Berge:2017ovy,Banerjee:2022sqg}, with low-energy neutrons \cite{Sponar:2020gfr} and gravitational wave detectors \cite{Vermeulen:2021epa}, by Big Bang Nucleosynthesis (BBN) \cite{Sibiryakov:2020eir,Bouley:2022eer,Ghosh:2025pbn} and observations of white dwarf mass-radius relations \cite{Bartnick:2025lbg}. Further tests have been proposed, including with space-based atomic clocks \cite{Brzeminski:2026rox}, atom interferometry \cite{MAGIS-100:2021etm,Bertoldi:2021rqk,AEDGE:2019nxb,Abend:2023jxv,Badurina:2019hst,Arvanitaki:2016fyj}, hyperfine transitions \cite{Hees:2016gop,Kennedy:2020bac},  molecular spectroscopy \cite{Oswald:2021vtc}, nuclear clocks \cite{Banks:2024sli} and pulsar timing arrays \cite{Gan:2025icr}.\footnote{We note that interactions between the scalar and matter can both enhance and suppress the sensitivity of experiments to the amplitude and time evolution of the background dark matter scalar \cite{Burrage:2025grx,Gan:2025nlu}.}

We assume the scalar to couple to the mass of Standard Model fermions, so that the classical dynamics of the scalar field are sourced by the trace of the matter stress-energy tensor, $T_\mu^\mu$. Such interactions can be generated, among others, from Higgs portal or conformal type couplings~\cite{Burrage:2018dvt,SevillanoMunoz:2024ayh}. When matter is relativistic, as in the early universe, one might thus expect the scalar field to be insensitive to matter fields with $T\approx 0$.  However, whenever a particle becomes non-relativistic, the trace of the matter stress-energy tensor becomes non-zero for around an e-fold, and `kicks' the scalar field. An effect that has been studied for dynamical dark energy models \cite{Erickcek:2013oma,Erickcek:2013dea,Padilla:2015wlv,Brax:2004qh} coupled scalar-tensor theories \cite{Damour:1993id}, and dilaton fields from string theory \cite{Damour:1994zq,Damour:1994ya}. This effect has also been seen previously in numerical simulations of quadratically-coupled ultra-light dark matter scalars in the early universe \cite{Sibiryakov:2020eir,Bouley:2022eer,Ghosh:2025pbn}, where constraints were placed on the theory by requiring that any variation in fundamental constants, caused by the evolution of the scalar, not disrupt BBN. In those works the effect of the kick from electrons becoming non-relativistic around the time of BBN was only determined numerically.  In what follows, we will show how  kicks from matter particles becoming non-relativistic enhance or suppress the initial scalar misalignment, depending on the strength and sign of the coupling, and show how the size of this effect can be estimated.

In Section \ref{sec: quadratic}, we introduce the scalar model we consider to describe ultra-light dark matter interacting quadratically with the Standard Model, and we determine how a matter particle becoming non-relativistic displaces the dark matter scalar. In Section \ref{sec:axion} we extend this to consider the potential and couplings of a dark axion model. We conclude in Section \ref{sec:conclusions}. 

\section{Quadratic model}\label{sec: quadratic}
In this section, we will introduce the effect that couplings between matter and scalar fields have on their evolution. For this, we will assume that we have an interaction of the form (mostly plus metric)
\begin{equation}
    \lgr\supset \frac{1}{2}(\nabla\phi)^2-V(\phi)+i\bar{\psi}_i\slashed{\nabla}\psi_i-m_i(1+A(\phi))\bar{\psi}_i\psi_i,
    \label{eq:lag}
\end{equation}
where $\phi$ is the scalar field that will behave as dark matter and $\psi_i$ is a matter field. 
In general, such an interaction would be expected to couple to more than one fermion species, depending on its nature. Here, however, we focus on the effect of a single coupling, noting that the analysis can be straightforwardly generalised to multiple particles \cite{Damour:2010rm, Damour:2010rp}. 
Assuming $|A(\phi)|<1$, we can approximate $m_i\bar{\psi_i}\psi_i\approx T{^\mu_\mu}_i$, with $T_{\mu\nu \,i}$ being the energy-momentum tensor for the fermion field. This leads to the following equation of motion for the $\phi$ field
\begin{equation}\label{eq: eom generic}
\begin{split}
    \ddot{\phi}+3H\dot{\phi}=&-V_{,\phi}(\phi)+A_{,\phi}(\phi)T{^\mu_\mu}_i,\\
    \approx&-V_{,\phi}(\phi)-A_{,\phi}(\phi)\rho_R \Sigma_i
\end{split}
\end{equation}
where $\rho_R$ is the energy density of the radiation bath, and we have assumed radiation domination to go from the first to the second line, allowing us to ignore non-relativistic degrees of freedom. $\Sigma_i=(\rho_i-3P_i)/\rho_R$ reads
\begin{equation}\label{eq: Sigma peak}
\edit{\Sigma_i(T)=\frac{15}{\pi^4}\frac{g_i}{g_*(T)}\left(\frac{m_i^2}{T^2}\right)\int^\infty_{m_i/T}\frac{\sqrt{u^2-(m_i/T)^2}}{e^u\pm 1}\d u,}
\end{equation}

\begin{figure}
\centering
\includegraphics[width=0.99\linewidth]{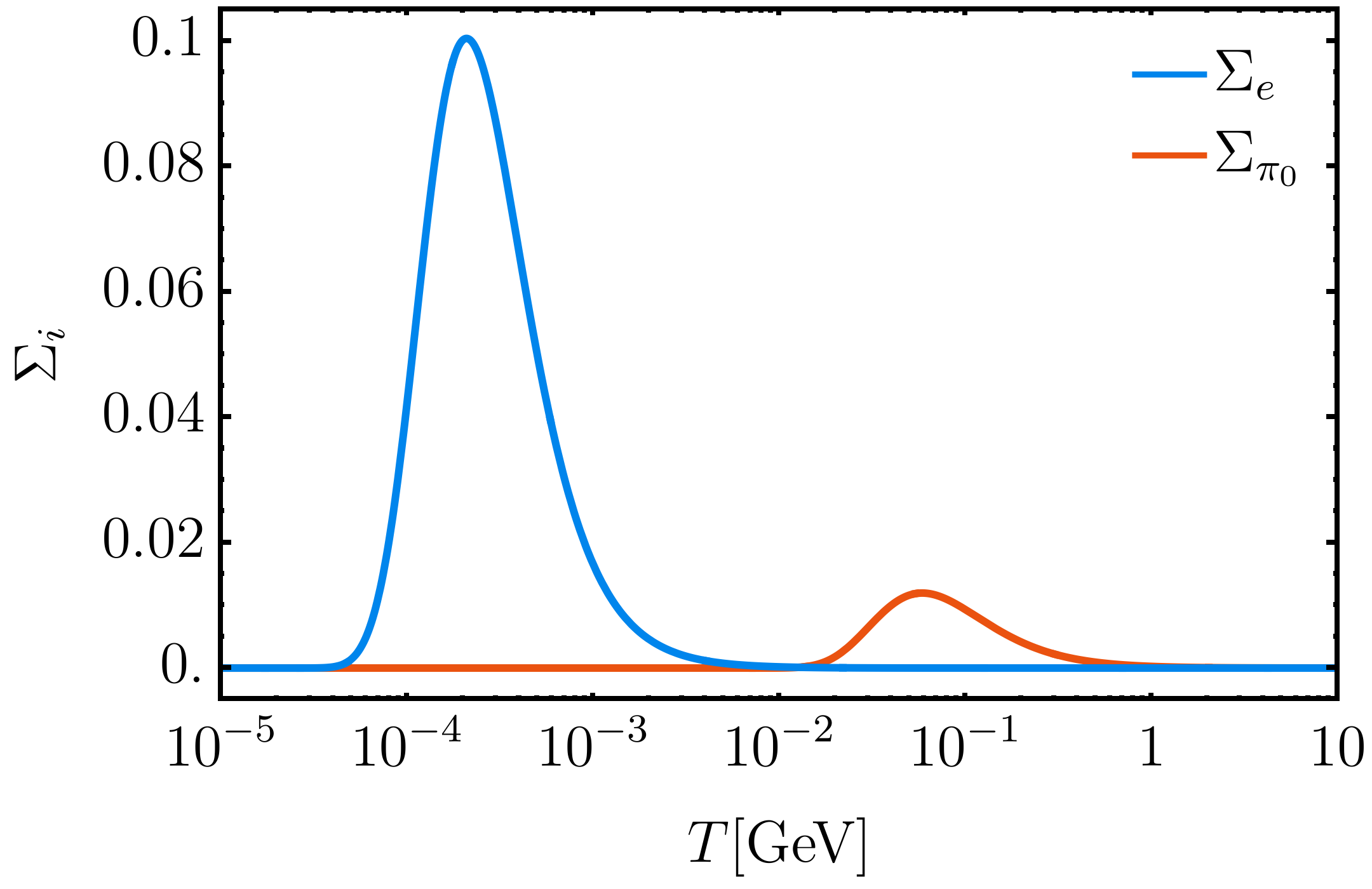}
\caption{\justifying Shape of the kick function $\Sigma_i(T)$ induced by coupling a scalar field to the trace of the energy-momentum tensor of a species $i$. \edit{We show the fermionic case for the electron, using the $+$ sign in Eq.~\eqref{eq: Sigma peak}, and the bosonic case for the neutral pion, using the $-$ sign. In both cases,} the kick peaks around $T\sim m_i$.
}
\label{fig:peak shape}
\end{figure}

\noindent \edit{where the $(+)$ sign applies to fermions and the $(-)$ sign to bosons} and $g_*(T)=\rho_R[(\pi^2/30)T^4]^{-1}$ is the number of relativistic degrees of freedom. This expression peaks for values around $T\sim m_i$, as shown in Figure~\ref{fig:peak shape} for the electron \edit{and neutral pion as representative fermionic and bosonic examples, respectively.}
 $\Sigma$ describes the influx of energy that goes from the fermion field into the $\phi$ field when $\psi_i$ becomes non-relativistic, and it's usually referred to as the ``kick"~\cite{Brax:2004qh}. While the effect of this kick has been studied in various scenarios, here we are interested in its impact on the initial conditions for ultra-light dark matter. 

Given that linear couplings of matter fields to a scalar field $\phi$ are very constrained due to fifth force experiments~\cite{Hees:2018fpg,Lee:2020zjt}, we will impose a $\mathbb{Z}_2$ symmetry on the model, leading to
\begin{align}
    V(\phi)=&\frac{\mu^2}{2}\phi^2, & A(\phi)=&\frac{\beta}{2\Mpl^2}\phi^2+ \cdots,
\end{align}
where the ellipsis contains higher order terms in $\beta \phi^2/\Mpl^2$. Therefore, to avoid any problems regarding the effective field theory, we will always ensure $A(\phi)<1$. Assuming a radiation-dominated universe ($\rho_R/H^2\approx 3\Mpl^2$), and thus $\rho_\phi\ll\rho_R$, this system will evolve under the following equation of motion
\begin{equation}
\begin{split}\label{eq:eom full}
   \phi''+\phi'\approx-\frac{\mu^2\phi}{H^2} - 3\beta \phi \Sigma(T_i),
\end{split}
\end{equation}
where the prime denotes the derivative with respect to e-folds, $N\equiv\log(a)$. Here, we focus on the effect of a single kick-type interaction, but one would expect one such kick per coupled fermion.

At early times, the scalar field dynamics are frozen due to the Hubble friction dominating over the influence of the potential, until the Hubble rate drops below the scalar mass, $H<m_\phi$. In what follows, we will analyse the effect of the fermion interaction on the dynamics of the $\phi$ field before unfreezing. The main difference to the standard evolution is that the unfreezing condition will now depend on the effective mass of the field, given by $m_{\rm eff}=\mu^2+3\beta H^2 \Sigma$. Since  we are interested in periods where $\mu<H$ but $m_{\rm eff}>H$, the equation of motion simplifies to
\begin{equation}
\begin{split}\label{eq: eom_final}
   \phi''+\phi' +3\beta \phi \Sigma(T_i)\approx0.
\end{split}
\end{equation}
Note that the Planck suppression in $A(\phi)$ is compensated by the large densities of the fields in the early universe to make the interaction term relevant. In addition, it is important to clarify that energy is conserved, the influx of energy into the scalar sector shifts  $\phi$, leading to a corresponding change in the mass of $\psi_i$. 

In what follows, we provide an analytical description of the scalar field’s evolution during the kick, which we will later validate with a numerical exploration of the parameter space. This provides an analytic guide to effects seen in  Refs.~\cite{Sibiryakov:2020eir, Bouley:2022eer} and extends the study to large couplings. \edit{We will focus on the effect of a fermionic kick, but the conclusion of this section can be easily extended to the bosonic case.} Since Eq.~\eqref{eq: eom_final} describes a simple harmonic oscillator, we will analyse the system’s behaviour separately in the oscillating and non-oscillating regimes:

\paragraph{\textbf{Oscillating:}}  In this limit, for a quadratic potential, once it unfreezes the scalar field starts oscillating around the minimum of its potential, behaving as pressureless matter. One could assume that the larger the interaction strength, the more energy is being injected into the scalar field by the kick function.
However, while the kick generates an instantaneous increase in the scalar potential energy of order $3\beta \Sigma \phi_i^2$, we also must consider that during the oscillations around the minimum the scalar energy density redshifts at a rate $\rho_\phi\propto a^{-3}$. This means that the field will lose a total energy of $\rho_\phi\approx e^{-3\Delta N_*}$ because of this interaction, where $\Delta N_*$ denotes the duration of the kick. These two effects compete with one another, and it is therefore not obvious, a priori, whether the oscillations persist long enough for the field to dissipate all of the energy injected by the kick.

To obtain the value for $\Delta N_*$, a first approximation is to determine the length of time that  $m_{\rm eff}>H$, or, assuming $m<H$, for how long $\sqrt{3}\beta \Sigma(N)>1$. Working with the kick equation from Eq.~\eqref{eq: Sigma peak}, we can approximate the growing and decaying parts of the kick function in Figure~\ref{fig:peak shape} respectively, by
    \begin{equation}
    \begin{split}
        \Sigma(T_+\gg m_i)&\approx m_i^2\,{ T_+^{-2}},\\
        \Sigma(T_-\ll m_i)&\approx m_i^{-5/2}T_-^{-5/2}e^{-m_i/T_-}.
        \end{split}
    \end{equation}
    From this, $\Delta N_*$ can be obtained by finding the root of $3\beta \Sigma(T)=1$ and then solving for $N$. For each part, the roots can be approximated by
    \begin{align}
        T_+=&m_i\sqrt{3\beta}\\
        T_-=& \frac{5}{2}\,m_iW_{-1}\left(\frac{2}{5}\beta^{2/5}\right)\approx m_i\frac{1}{\log(3\beta)+\frac{5}{2}\log(\log(3\beta))}, \nonumber
    \end{align}
    where $W_{-1}(x)$ is the $k=-1$ branch of the Lambert W function, that solves for the root of the decaying part of the kick function, for which the approximation holds as long as $\beta\gg1$. $\Delta N_*$ is related to these temperatures by
    \begin{equation}
        \Delta N_*=\log\left(\frac{T_+}{T_-}\right),
    \end{equation}
    from which we obtain
    \begin{equation}\label{eq:DeltaN*}
        \Delta N_*=\log{\left(\sqrt{3\beta}\left(\log(3\beta)+\frac{5}{2}\log(\log(3\beta))\right)\right)}.
    \end{equation}

    Therefore, assuming that the kick gives all its energy to the scalar field and the field oscillates for a period of $\Delta N_*$, its energy density by the end of the kick will be
    \begin{equation}
    \langle\rho_\phi\rangle\propto\frac{3\phi_i^2}{2}\beta \Sigma|_{\rm max} \exp\left(-\frac{3}{2}\Delta N_*\right),
    \end{equation}
    where $ \Sigma|_{\rm max}$ is the maximum value the kick function will take at its peak, which is around $0.1\beta$. Substituting for Eq.~\eqref{eq:DeltaN*} and taking the term at leading order in $\beta$, we find 
    \begin{equation}\label{eq: rho_phi sqrt beta}
    \langle\rho_\phi\rangle\propto\beta \exp\left(-\frac32\log(\sqrt{\beta})\right)=\frac{1}{\sqrt{\beta}}.
    \end{equation}
    Therefore, for large values of $\beta$, the total energy of the scalar field will always decrease due to the kick.

\begin{figure*}
    \centering
    \includegraphics[width=0.99\linewidth]{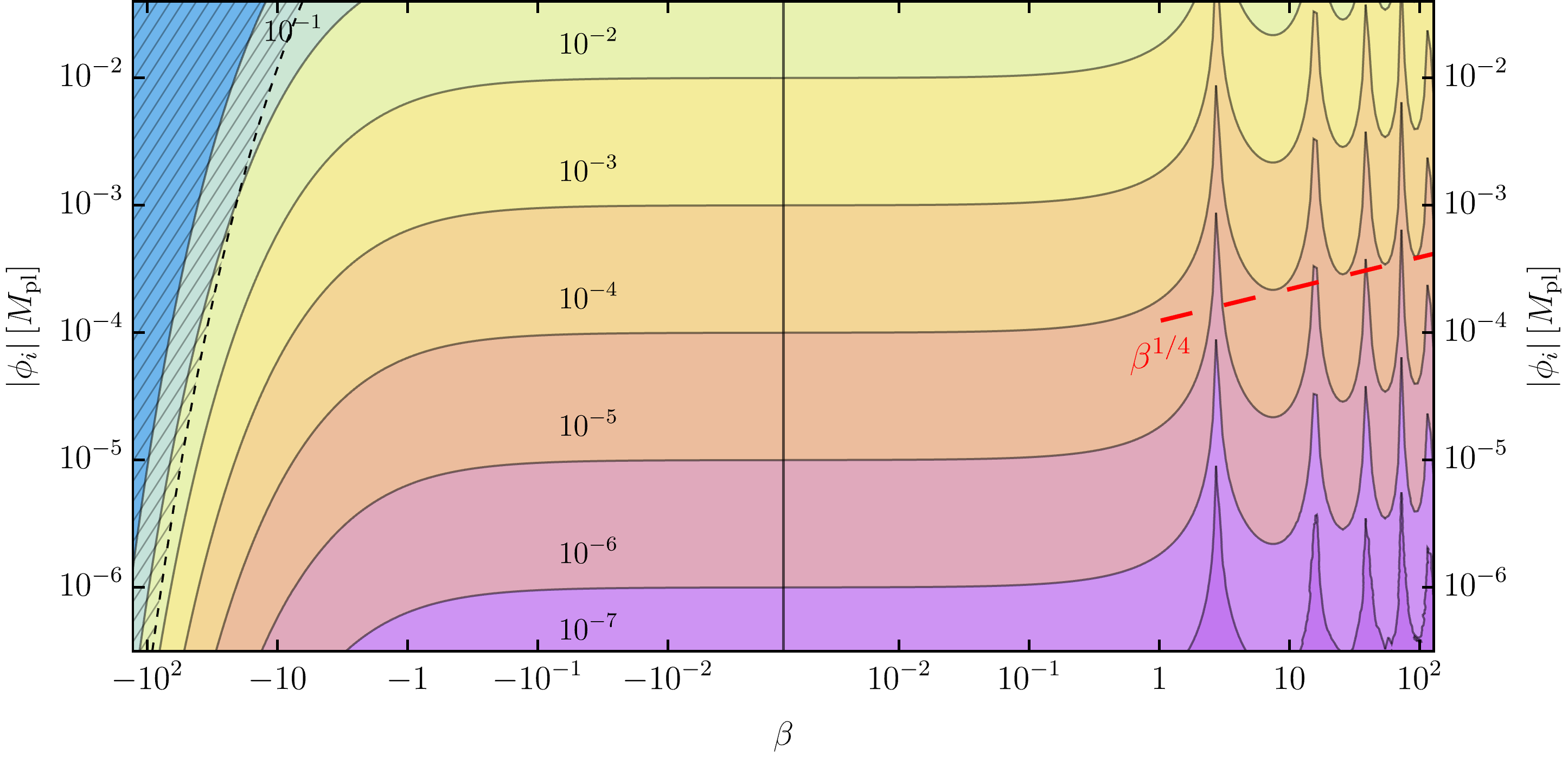}
    \caption{\justifying Final value of the scalar fields as a function of the initial value $|\phi_i|$ and both negative and positive kick strengths, $\beta$, after numerical evolution of the system. The different coloured regions show the order of magnitude of  the value of the field at the end of the kick ($|\phi(N_f)|$). Dark blue on the left marks the region where $\phi>M_{\rm pl}$, and the dashes mark the region where $|A(\phi_f)|>1$. The rightmost peaks in the figure mark scenarios where, at the end of the evolution, the field is in the  minimum of the quadratic potential.}
    \label{fig:numerics}
\end{figure*}

\paragraph{\textbf{Non-oscillating:}} This occurs when the interaction strength (i.e., $\beta \Sigma$) is very small or negative. If $\Sigma(T)$ were constant, an analytical solution from Eq.~\eqref{eq: eom_final} would straightforwardly follow using the ansatz $\phi\propto e^{\lambda N}$. In our case, however, we only assume that $\Sigma(T)$ varies slowly, which is not always guaranteed. Nevertheless, we can average the effect of its evolution on $\phi$ and obtain a good approximate solution:
\begin{equation}
    \phi(N)=\phi_i\exp\left(\int^N_{N_i}\frac{-1+\sqrt{1-12\beta\Sigma(\hat{N})}}{2} {d} \hat{N}\right),
\end{equation}
which, depending on the size of $\beta$, we can approximate as
\begin{equation}\label{eq:phi_evol}
 \phi_f\approx\phi_i
\begin{cases}
   \exp\left(-\beta\bar{\Sigma}\right) &\text{ for  } |\beta|<1\\\\
   \exp\left(\sqrt{-\beta\bar{\Sigma}}\right) &\text{ for } \beta\ll-1,
\end{cases}    
\end{equation}
where $\bar{\Sigma}$ is the average value for $\Sigma(N)$.

We compare this analytical approach against numerical computations in Figure~\ref{fig:numerics}, both for positive and negative interaction strengths. For large positive kicks, the field loses energy due to the extra oscillations it undergoes, with this effect peaking every time the field  ends its evolution  near the minimum of the potential. However, while the $\beta$ dependence is highly oscillatory, we can still see the overall trend of $\phi_i\propto\beta^{1/4}$ from Eq.~\eqref{eq: rho_phi sqrt beta} marked by the dashed red line. In the negative kick case, we find that, at large $\beta$, the field is very sensitive to the instability created by flipping the sign of the quadratic potential. In summary, the overall dependence of the strength of quadratic couplings on the energy of a scalar field is
\begin{equation}
     \langle\rho_f\rangle\propto
\begin{cases}
    \beta^{-1/2} &\text{ for  } \beta\gg1\\\\
   \exp\left(-\beta\bar{\Sigma}\right) &\text{ for  } |\beta|<1\\\\
   \exp\left(\sqrt{-\beta\bar{\Sigma}}\right) &\text{ for } \beta\ll-1,
\end{cases}    
\end{equation}

We can see that even Planck-suppressed interactions can have a large effect on the evolution of scalar fields in the early universe, independently of the Hubble parameter at the time. For the quadratic model presented in this section, we must remain within the regime of validity of the EFT (i.e., $|A(\phi)|<1$ at all times). This is especially challenging for large negative $\beta$ since the instability in the potential tends to displace the field over this limit, as marked by the black dashed region in Figure~\ref{fig:numerics}. We will now show how our conclusions differ for an axion-like potential.

\section{Dark QCD Axion model}
\label{sec:axion}
Axion fields are among the most compelling dark-matter candidates due to their close connection to fundamental physics. Here we consider the often studied possibility that axion dark matter is part of a “dark QCD” sector~\cite{Dondi:2019olm,Garani:2021zrr,Tsai:2020vpi,Morrison:2020yeg,Khoury:2025txd,Hook:2017psm}. In that case, generic interactions among the dark-sector fields are expected and, because the axion is a pseudo-scalar field, interactions with, for example, fermion mass terms appear first at order $\phi^2$,  leading to similar dynamics to those discussed above. Throughout this section, QCD terminology (e.g., quarks, baryons, and hadronic parameters) refers to the dark sector unless explicitly labeled as Standard Model QCD (SM-QCD).

The bare axion potential due to the pion mass terms~\cite{DiVecchia:1980yfw,GrillidiCortona:2015jxo} is given by
\begin{equation}\label{eq: free V axion}
    V(a)=-\Lambda^4 \sqrt{1-\xi\sin^2\left(\frac{a}{2f}\right)},
\end{equation}
where we take $\xi=4\frac{m_u m_d}{(m_u+m_d)^2}\approx0.86$ to be the dimensionless ratio of quark masses to mimic the SM-QCD axion behaviour, and $\Lambda^4=\epsilon \,m_\pi^2f_\pi^2$ is an energy scale related to the mass and decay constant of the neutral pion, respectively. In this way, the axion mass is given by $m_a=\frac{\Lambda^2\sqrt{\xi}}{2f}$. 
Following \cite{Hook:2017psm,Khoury:2025txd}, we have introduced the parameter $\epsilon\ll1$ that allows for finite density corrections, which we will introduce later, to overcome the vacuum potential in Eq.~\eqref{eq: free V axion} without leaving the perturbative regime. 

In the presence of a background density of dark baryons, the axion picks up an effective correction to the potential that can be generally expressed as
\begin{equation}
\begin{split}
     V_b(a)=+&2\alpha \sqrt{1-\xi\sin^2\left(\frac{a}{2f}\right)}{T^\mu_\mu}^{(b)}\\+&\mathcal{O}\left(\left(\frac{\alpha}{m_\pi^2f_\pi^2} T{^\mu_\mu}^{(b)}\right)^2\right),
\end{split}
\end{equation}
where $\alpha$ controls the strength of the interaction and $T{^\mu_\mu}^{(b)}$ is the energy density of the dark baryon species coupling to the axion field. For reference, $\alpha$ is fixed to $\alpha={\sigma_N}/{m_N} $ for \edit{nucleon} interactions in the SM-QCD axion, with $\sigma_N=59\pm7\,\text{ MeV}$ being the pion-nucleon sigma term and $m_N$ the average mass of nucleons. \edit{One can also consider corrections from lighter states such as the neutral pion itself, for which the leading order correction corresponds to taking $\alpha=1/2$~\cite{GrillidiCortona:2015jxo,Alarcon:2011zs}.} In total, the evolution of this field is dictated by the effective potential
\begin{equation}
\begin{split}
    V_{eff}(a)&=-
    \Lambda^4\left(1-\frac{2\alpha }{\Lambda^4}T{^\mu_\mu}^{(b)}\right)\sqrt{1-\xi\sin^2\left(\frac{a}{2f}\right)}\\&+\mathcal{O}\left(\left(\frac{\alpha}{m_\pi^2f_\pi^2} T{^\mu_\mu}^{(b)}\right)^2\right),
\end{split}
\end{equation}

\begin{figure}
    \centering
    \includegraphics[width=1\linewidth]{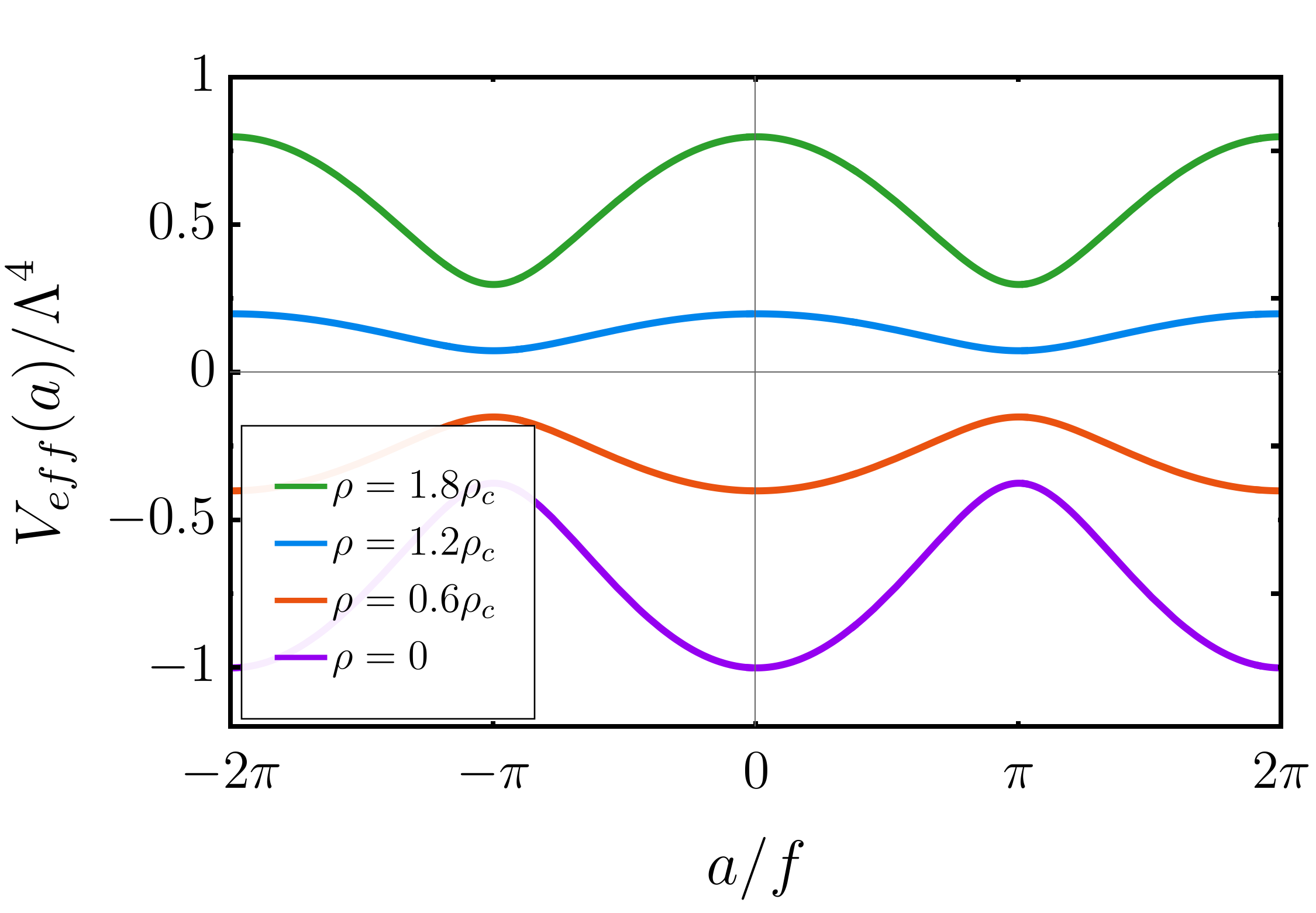}
    \caption{\justifying Effective potential for the axion model when including interactions with \edit{dark baryons}. The density corrections lead to a sign swap in the early universe for densities larger than $\rho_c\equiv\Lambda^4/2\alpha$.}
    \label{fig: Veff axion}
\end{figure}

\noindent  As shown in Figure~\ref{fig: Veff axion}, large density corrections can flip the sign of the effective potential. This is the key aspect of the dark QCD axion, as when a dark baryon species becomes non-relativistic in the thermal bath, it will displace the field to the maximum of the bare potential. For this to take place, we need to satisfy the following two conditions:
\begin{itemize}
     \item[I.] \textit{Perturbative control:}
        \begin{equation}
            \frac{\alpha }{m_\pi^2f_\pi^2}T{^\mu_\mu}^{(b)}<1.
        \end{equation}
        
    \item[II.] \textit{Sign-switch on effective potential:}
        \begin{equation}
            \frac{2\alpha }{\Lambda^4}T{^\mu_\mu}^{(b)}>1 ,
        \end{equation}
\end{itemize}

For these conditions to be satisfied, we only need to impose $\varepsilon\ll1$, which can be achieved by having multi-axion theories with $\mathbb{Z}_N$ exchange symmetry~\cite{Hook:2018jle,Alarcon:2011zs}. Additionally, assuming dark baryons exist as a stable species before they become non-relativistic requires the axion scale to be larger than the energy density of baryons, meaning $ T{^\mu_\mu}^{(b)}<\edit{m_\pi^2f_\pi^2}$. This allows us to use the well-known formula for the kick (Eq.~\eqref{eq: Sigma peak}) to describe the time evolution of this interaction.\footnote{ If the dark baryons are formed already non-relativistic, as is the case for \edit{nucleons in QCD without further model-building}, they will also provide a kick-like interaction into the dynamics of the axion field \edit{at the QCD phase transition}. However, its shape and strength is difficult to estimate \edit{as it requires including non-perturbative terms}, and best attempts involve complicated numerical analyses~\cite{Saikawa:2018rcs}.} 

\begin{figure}
    \includegraphics[width=\linewidth]{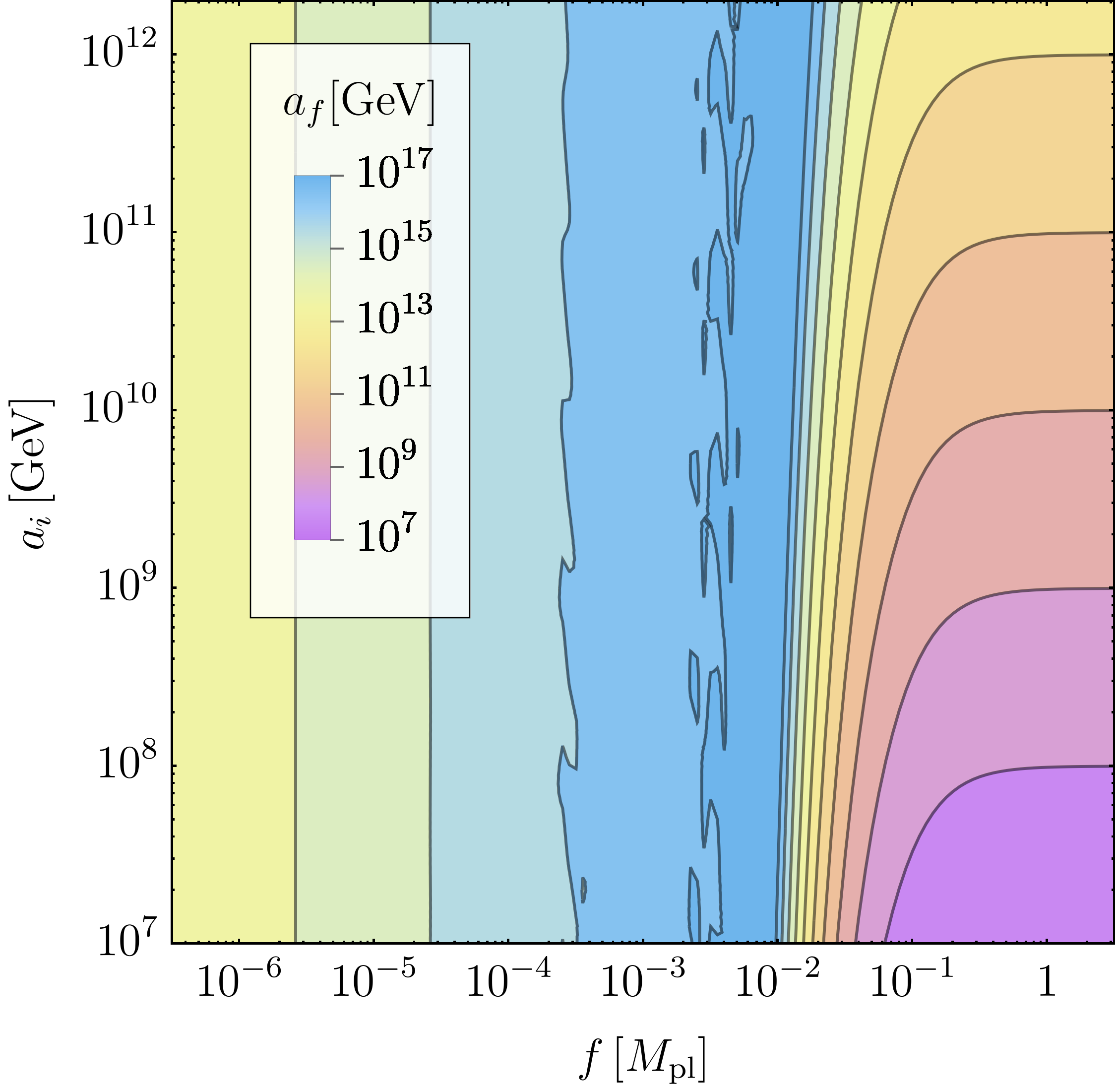}
    \caption{\justifying Final value of the axion field at the end of the kick due to interactions with \edit{pions} for different \edit{decay constants} and initial values. For this figure, we fixed $\alpha= \edit{1/2}$ as given \edit{for the pion interaction}. The different coloured regions indicate the order of magnitude of the value of the field at the end of the kick ($a(N_f)$). For large masses, the field reaches the effective minimum placed at $a_f=\pi f$. }
    \label{fig:axion}
\end{figure}

Assuming the universe to be radiation dominated, the equation of motion will take the following form
\begin{equation}
\begin{split}
    a''+a'\approx&-\frac{\sin\left(\frac{a}{2f}\right)\cos\left(\frac{a}{2f}\right)}{\sqrt{1-\xi\sin^2\left(\frac{a}{2f}\right)}}\left(\frac{2m_a^2 f}{H^2}-\beta_b \pi f\Sigma^{(b)}\right),
    \end{split}
\end{equation}
where $\beta_b=\frac{3\Mpl^2\alpha\xi}{\pi f^2}$ is the interaction strength and $\Sigma^{(b)}$ is the kick function associated with the dark baryon using Eq.~\eqref{eq: Sigma peak}, which describes the energy influx to the axion field when the baryons go from being relativistic to non-relativistic. As in Section~\ref{sec: quadratic}, we will assume the bare mass of the potential to be smaller than the Hubble constant, such that the field unfreezes because of the interaction to matter. In Figure~\ref{fig:axion} we show the field rescaling from $a_i$ to $a_f$ due to the \edit{pion} kick, \edit{meaning $\alpha=1/2$}, for a dark QCD axion potential that shares the same $m_{\pi}$ as in SM-QCD. For $f>10^{-2}\Mpl$, we find a similar result to that for negative $\beta$ in the quadratic model, shown in Figure~\ref{fig:numerics}. However, for smaller \edit{decay constants} the interaction becomes stronger, and the field is able to reach the minimum of the effective potential, located at $a_{max}=\pi f$, regardless of the starting position. \edit{Between these two regimes, the kick does not last long enough for the field to lose all of its momentum and settle at the effective minimum, leading to a non-trivial pattern in the final value of the axion field.}

\begin{figure}
    \centering
    \includegraphics[width= \linewidth]{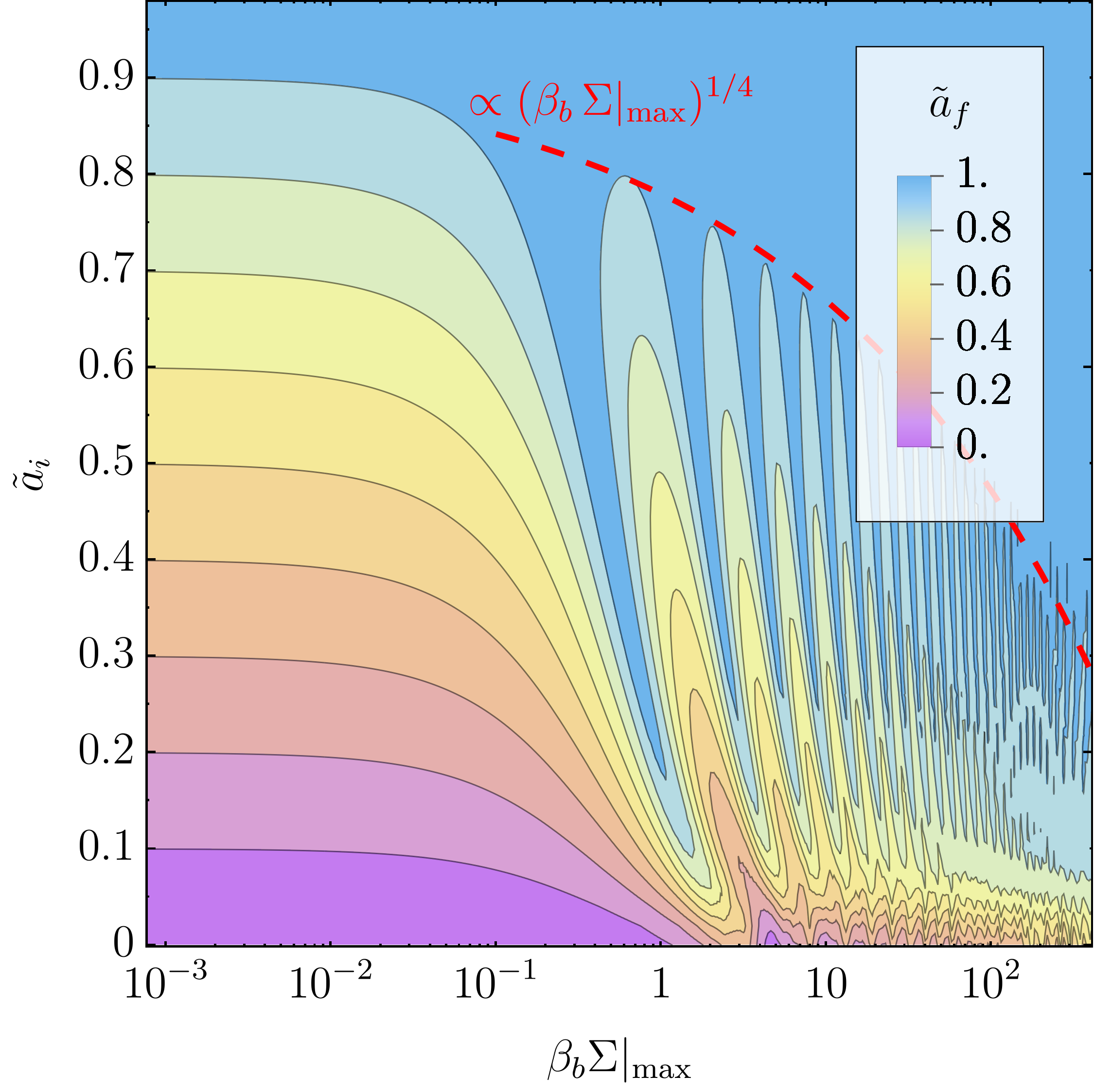}
    \caption{ \justifying Final value of the rescaled axion field,  $\tilde{a}=a\pi f$, at the end of the kick for different coupling $\beta_b=\frac{3\Mpl^2\alpha\xi}{f^2}$, and initial conditions, $\tilde{a}_i$, given some kick peak value $\Sigma|_{\rm max}$. The different coloured regions indicate the order of magnitude of the field at the end of the kick ($\tilde{a}_f$), where the same color is given for symmetric points with respect to the maximum $\tilde{a}_f=1$.}
    \label{fig: tildeaxion}
\end{figure}

Although this system has multiple free parameters, we can simplify the full model by rescaling. $a\to f\pi\tilde{a}$.\footnote{For this rescaling to be valid, we must ensure that the universe stays radiation dominated. In general, this requires $\tilde a'\pi f \ll M_{\rm Pl}$, which is satisfied provided $f\ll M_{\rm Pl}$} This leads to the following equation of motion
\begin{equation}
     \tilde{a}''+ \tilde{a}'\approx-\frac{\sin\left(\frac{ \tilde{a}\pi}{2}\right)\cos\left(\frac{ \tilde{a}\pi}{2}\right)}{\sqrt{1-\xi\sin^2\left(\frac{ \tilde{a}\pi}{2}\right)}}\left(\frac{2\pi m_a^2}{H^2}-\beta_b \Sigma^{(b)}\right),
\end{equation}
The behaviour of the generic system is illustrated in Figure~\ref{fig: tildeaxion}, which shows the mapping of initial values $\tilde a_i$ to final values $\tilde a_f$ for different interaction strengths $\beta_b$. As expected, for $\beta_b<10$ the dynamics closely resemble the negative-$\beta$ quadratic case. For stronger interactions, we instead observe oscillations about the new minimum at $\tilde a = 1$.


Although an analytic treatment is challenging due to the nonlinearity of both the potential and the coupling function, we can still use the approximation from the previous section in the regime of small amplitudes of oscillation. This corresponds to initial conditions sufficiently close to the effective minimum, $\tilde a = 1$. In that limit, the scaling $\tilde a_f \propto 1-(\beta_b\Sigma|_{\rm max})^{1/4}$ provides a good description of the oscillation envelope, as marked in Figure~\ref{fig: tildeaxion} by the dashed red line for $\tilde a_f \simeq 0.9$. For smaller $\tilde a_f$, this approximation breaks down due to the quadratic expansion not being a good approximation for the full axion potential.

The calculations shown in this section illustrate the positive aspects of these interactions on the axion potential, as they can alleviate the fine-tuning issues associated with the misalignment mechanism. This is because, independently of their initial position, interactions controlled by scales a couple of orders of magnitude larger than the Planck scale will drive the field all the way to the top of the axion potential.

\section{Conclusions}
\label{sec:conclusions}
In this work, we have examined how particle interactions affect the early-time dynamics of ultra-light dark matter. When a particle species becomes non-relativistic, it induces a “kick”-like effect on the scalar field, temporarily destabilizing it. We have studied this effect in the context of the misalignment mechanism, a key ingredient in ultra-light dark-matter models, in which the field must begin displaced from its minimum so that the amplitude of its oscillations carries sufficient energy density today.

We first considered a quadratic potential for the dark matter field, with a quadratic coupling to matter. By analytically estimating the duration of the kick, we studied the effect of this interaction for different coupling strengths, $\beta$. For large positive couplings, the kick induces a large effective mass on the scalar field, making it oscillate around the minimum of the potential before it would otherwise. Overall, these oscillations cause the total energy of the field to reduce by $\rho_f\propto\beta^{-1/2}$. On the contrary, for large negative couplings, the quadratic potential flips its sign and the total energy is increased by $\log(\rho_f)\propto(-\beta\bar{\Sigma})^{1/2}$, where $\bar{\Sigma}$ is the average amplitude of the kick function as given in Eq.~\eqref{eq: Sigma peak}. We provide a numerical exploration of the parameter space for the quadratic model in Figure~\ref{fig:numerics}.

We next applied the same analysis to a dark-QCD axion. In the small amplitude regime, the dynamics reduce to those of a quadratic potential, but the cosine-like structure bounds the field displacement for large interactions. Because the coupling to dark baryons is negative, the kick can reverse the sign of the effective potential, placing its minimum at the top of the bare potential during the kick. While the full system can not be easily described analytically, the quadratic analysis remains applicable for small oscillations about the effective minimum. The results for a generic potential and different coupling strengths are shown in Figure~\ref{fig: tildeaxion}. We also considered the evolution for an axion with the Standard Model's QCD parameter choices in Figure~\ref{fig:axion}. We find that in the weak-coupling limit, the behaviour closely tracks the quadratic case, while for sufficiently strong couplings, the field relaxes to an approximately constant value set by the new minimum of the effective potential.

In conclusion, particle interactions can significantly affect the early evolution of ultra-light dark matter, even when the coupling strength is only a few orders of magnitude larger than gravity. The effect depends on the sign of the coupling, which may either increase or decrease the initial vacuum expectation value of the scalar field, thereby improving or worsening the misalignment needed to account for the observed dark matter abundance today. For a dark-QCD axion, as the sign of the coupling is fixed, we find that particle interactions always enhance the late-time dark matter abundance.

\section*{Acknowledgments}
The authors thank Joerg Jaeckel, Adam Moss, and Xavier Pritchard for useful discussions. CB would like to thank Christian Byrnes, Swagat Saurav Mishra,  Xavier Pritchard, and David Seery for collaboration on related work. CB is supported by the STFC under Grant No. ST/T000732/1. SSM is supported by the STFC under Grant No.~ST/T001011/1 and by funds provided by the Center for Particle
Cosmology at the University of Pennsylvania.

\section*{Data Availability}

The Wolfram Mathematica code used to generate all figures in this work is publicly available in Ref.~\cite{SevillanoMunoz:Code}.

\bibliographystyle{BiblioStyle}
\bibliography{DraftBiblio}

@article{Bouley:2022eer,
    author = "Bouley, Thomas and S{\o}rensen, Philip and Yu, Tien-Tien",
    title = "{Constraints on ultralight scalar dark matter with quadratic couplings}",
    eprint = "2211.09826",
    archivePrefix = "arXiv",
    primaryClass = "hep-ph",
    reportNumber = "DESY-22-186",
    doi = "10.1007/JHEP03(2023)104",
    journal = "JHEP",
    volume = "03",
    pages = "104",
    year = "2023"
}

@article{Ghosh:2025pbn,
    author = "Ghosh, Subhajit and Boddy, Kimberly K. and Yu, Tien-Tien",
    title = "{Early Universe Constraints on Variations in Fundamental Constants Induced by Ultralight Scalar Dark Matter}",
    eprint = "2511.14532",
    archivePrefix = "arXiv",
    primaryClass = "astro-ph.CO",
    reportNumber = "UT-WI-39-2025",
    month = "11",
    year = "2025"
}

@article{Brzeminski:2026rox,
    author = "Brzeminski, Dawid and Pierce, Aaron",
    title = "{Searching for Ultralight Scalar Dark Matter with Clocks in Low Earth Orbit}",
    eprint = "2601.16259",
    archivePrefix = "arXiv",
    primaryClass = "hep-ph",
    reportNumber = "LITP-26-03",
    month = "1",
    year = "2026"
}

@article{Banerjee:2022sqg,
    author = "Banerjee, Abhishek and Perez, Gilad and Safronova, Marianna and Savoray, Inbar and Shalit, Aviv",
    title = "{The phenomenology of quadratically coupled ultra light dark matter}",
    eprint = "2211.05174",
    archivePrefix = "arXiv",
    primaryClass = "hep-ph",
    doi = "10.1007/JHEP10(2023)042",
    journal = "JHEP",
    volume = "10",
    pages = "042",
    year = "2023"
}

@article{Gan:2025icr,
    author = "Gan, Xucheng and Kim, Hyungjin and Mitridate, Andrea",
    title = "{Probing Quadratically Coupled Ultralight Dark Matter with Pulsar Timing Arrays}",
    eprint = "2510.13945",
    archivePrefix = "arXiv",
    primaryClass = "hep-ph",
    reportNumber = "DESY-25-134",
    month = "10",
    year = "2025"
}

@article{Bartnick:2025lbg,
    author = "Bartnick, Kai and Springmann, Konstantin and Stelzl, Stefan and Weiler, Andreas",
    title = "{$\phi$-Dwarfs: White Dwarfs probe Quadratically Coupled Scalars}",
    eprint = "2509.25305",
    archivePrefix = "arXiv",
    primaryClass = "hep-ph",
    reportNumber = "TUM-HEP-1569/25",
    month = "9",
    year = "2025"
}

@article{Burrage:2025grx,
    author = "Burrage, Clare and Macdonald, Angus and Ross, Michael P. and Rybka, Gray and Todarello, Elisa",
    title = "{Impact of cavities on the detection of quadratically coupled ultralight dark matter}",
    eprint = "2507.16526",
    archivePrefix = "arXiv",
    primaryClass = "hep-ph",
    doi = "10.1103/bml1-j932",
    journal = "Phys. Rev. D",
    volume = "113",
    number = "6",
    pages = "063505",
    year = "2026"
}

@article{Gan:2025nlu,
    author = "Gan, Xucheng and Liu, Da and Liu, Di and Luo, Xuheng and Yu, Bingrong",
    title = "{Detecting ultralight dark matter with matter effect}",
    eprint = "2504.11522",
    archivePrefix = "arXiv",
    primaryClass = "hep-ph",
    reportNumber = "DESY-25-060",
    doi = "10.1007/JHEP02(2026)043",
    journal = "JHEP",
    volume = "02",
    pages = "043",
    year = "2026"
}

@article{Kim:2022ype,
    author = "Kim, Hyungjin and Perez, Gilad",
    title = "{Oscillations of atomic energy levels induced by QCD axion dark matter}",
    eprint = "2205.12988",
    archivePrefix = "arXiv",
    primaryClass = "hep-ph",
    reportNumber = "DESY-22-088",
    doi = "10.1103/PhysRevD.109.015005",
    journal = "Phys. Rev. D",
    volume = "109",
    number = "1",
    pages = "015005",
    year = "2024"
}

@article{Panda:2023nir,
    author = {Panda, Cristian D. and Tao, Matthew J. and Ceja, Miguel and Khoury, Justin and Tino, Guglielmo M. and M{\"u}ller, Holger},
    title = "{Measuring gravitational attraction with a lattice atom interferometer}",
    eprint = "2310.01344",
    archivePrefix = "arXiv",
    primaryClass = "physics.atom-ph",
    doi = "10.1038/s41586-024-07561-3",
    journal = "Nature",
    volume = "631",
    number = "8021",
    pages = "515--520",
    year = "2024"
}

@article{Filzinger:2023qqh,
    author = "Filzinger, Melina and Caddell, Ashlee R. and Jani, Dhruv and Steinel, Martin and Giani, Leonardo and Huntemann, Nils and Roberts, Benjamin M.",
    title = "{Ultralight Dark Matter Search with Space-Time Separated Atomic Clocks and Cavities}",
    eprint = "2312.13723",
    archivePrefix = "arXiv",
    primaryClass = "hep-ph",
    doi = "10.1103/PhysRevLett.134.031001",
    journal = "Phys. Rev. Lett.",
    volume = "134",
    number = "3",
    pages = "031001",
    year = "2025"
}

@article{Arvanitaki:2016fyj,
    author = "Arvanitaki, Asimina and Graham, Peter W. and Hogan, Jason M. and Rajendran, Surjeet and Van Tilburg, Ken",
    title = "{Search for light scalar dark matter with atomic gravitational wave detectors}",
    eprint = "1606.04541",
    archivePrefix = "arXiv",
    primaryClass = "hep-ph",
    doi = "10.1103/PhysRevD.97.075020",
    journal = "Phys. Rev. D",
    volume = "97",
    number = "7",
    pages = "075020",
    year = "2018"
}

@article{Sponar:2020gfr,
    author = "Sponar, Stephan and Sedmik, Ren{\'e} I. P. and Pitschmann, Mario and Abele, Hartmut and Hasegawa, Yuji",
    title = "{Tests of fundamental quantum mechanics and dark interactions with low-energy neutrons}",
    eprint = "2012.09048",
    archivePrefix = "arXiv",
    primaryClass = "quant-ph",
    doi = "10.1038/s42254-021-00298-2",
    journal = "Nature Rev. Phys.",
    volume = "3",
    number = "5",
    pages = "309--327",
    year = "2021"
}

@article{Elder:2025tue,
    author = "Elder, Benjamin and Mentasti, Giorgio and Pasatembou, Elizabeth and Baynham, Charles F. A. and Buchmueller, Oliver and Contaldi, Carlo R. and de Rham, Claudia and Hobson, Richard and Tolley, Andrew J.",
    title = "{Prospects for detecting new dark physics with the next generation of atomic clocks}",
    eprint = "2504.07179",
    archivePrefix = "arXiv",
    primaryClass = "hep-ph",
    reportNumber = "Imperial/TP/2025/CC/1, Imperial--TP--2025--CC--1",
    doi = "10.1103/vghm-4s5p",
    journal = "Phys. Rev. D",
    volume = "112",
    number = "4",
    pages = "044053",
    year = "2025"
}

@article{Vermeulen:2021epa,
    author = "Vermeulen, Sander M. and others",
    title = "{Direct limits for scalar field dark matter from a gravitational-wave detector}",
    eprint = "2103.03783",
    archivePrefix = "arXiv",
    primaryClass = "gr-qc",
    doi = "10.1038/s41586-021-04031-y",
    month = "3",
    year = "2021"
}

@article{Berge:2017ovy,
    author = "Berg{\'e}, Joel and Brax, Philippe and M{\'e}tris, Gilles and Pernot-Borr{\`a}s, Martin and Touboul, Pierre and Uzan, Jean-Philippe",
    title = "{MICROSCOPE Mission: First Constraints on the Violation of the Weak Equivalence Principle by a Light Scalar Dilaton}",
    eprint = "1712.00483",
    archivePrefix = "arXiv",
    primaryClass = "gr-qc",
    doi = "10.1103/PhysRevLett.120.141101",
    journal = "Phys. Rev. Lett.",
    volume = "120",
    number = "14",
    pages = "141101",
    year = "2018"
}

@article{MAGIS-100:2021etm,
    author = "Abe, Mahiro and others",
    collaboration = "MAGIS-100",
    title = "{Matter-wave Atomic Gradiometer Interferometric Sensor (MAGIS-100)}",
    eprint = "2104.02835",
    archivePrefix = "arXiv",
    primaryClass = "physics.atom-ph",
    reportNumber = "FERMILAB-PUB-21-031-AD-DI-FESS-QIS-T",
    doi = "10.1088/2058-9565/abf719",
    journal = "Quantum Sci. Technol.",
    volume = "6",
    number = "4",
    pages = "044003",
    year = "2021"
}

@article{Bertoldi:2021rqk,
    author = "Bertoldi, Andrea and others",
    title = "{AEDGE: Atomic experiment for dark matter and gravity exploration in space}",
    doi = "10.1007/s10686-021-09701-3",
    journal = "Exper. Astron.",
    volume = "51",
    number = "3",
    pages = "1417--1426",
    year = "2021"
}

@article{AEDGE:2019nxb,
    author = "El-Neaj, Yousef Abou and others",
    collaboration = "AEDGE",
    title = "{AEDGE: Atomic Experiment for Dark Matter and Gravity Exploration in Space}",
    eprint = "1908.00802",
    archivePrefix = "arXiv",
    primaryClass = "gr-qc",
    reportNumber = "KCL-PH-TH/2019-65, CERN-TH-2019-126",
    doi = "10.1140/epjqt/s40507-020-0080-0",
    journal = "EPJ Quant. Technol.",
    volume = "7",
    pages = "6",
    year = "2020"
}

@article{Abend:2023jxv,
    author = "Abend, Sven and others",
    title = "{Terrestrial Very-Long-Baseline Atom Interferometry: Workshop Summary}",
    eprint = "2310.08183",
    archivePrefix = "arXiv",
    primaryClass = "hep-ex",
    reportNumber = "CERN-TH-2023-176",
    doi = "10.1116/5.0185291",
    journal = "AVS Quantum Sci.",
    volume = "6",
    number = "2",
    pages = "024701",
    year = "2024"
}

@article{Badurina:2019hst,
    author = "Badurina, L. and others",
    title = "{AION: An Atom Interferometer Observatory and Network}",
    eprint = "1911.11755",
    archivePrefix = "arXiv",
    primaryClass = "astro-ph.CO",
    reportNumber = "AION-2019-001, CERN-TH-2019-199",
    doi = "10.1088/1475-7516/2020/05/011",
    journal = "JCAP",
    volume = "05",
    pages = "011",
    year = "2020"
}

@article{Hees:2016gop,
    author = "Hees, A. and Gu{\'e}na, J. and Abgrall, M. and Bize, S. and Wolf, P.",
    title = "{Searching for an oscillating massive scalar field as a dark matter candidate using atomic hyperfine frequency comparisons}",
    eprint = "1604.08514",
    archivePrefix = "arXiv",
    primaryClass = "gr-qc",
    doi = "10.1103/PhysRevLett.117.061301",
    journal = "Phys. Rev. Lett.",
    volume = "117",
    number = "6",
    pages = "061301",
    year = "2016"
}

@article{Kennedy:2020bac,
    author = "Kennedy, Colin J. and Oelker, Eric and Robinson, John M. and Bothwell, Tobias and Kedar, Dhruv and Milner, William R. and Marti, G. Edward and Derevianko, Andrei and Ye, Jun",
    title = "{Precision Metrology Meets Cosmology: Improved Constraints on Ultralight Dark Matter from Atom-Cavity Frequency Comparisons}",
    eprint = "2008.08773",
    archivePrefix = "arXiv",
    primaryClass = "physics.atom-ph",
    doi = "10.1103/PhysRevLett.125.201302",
    journal = "Phys. Rev. Lett.",
    volume = "125",
    number = "20",
    pages = "201302",
    year = "2020"
}

@article{Oswald:2021vtc,
    author = "Oswald, R. and others",
    title = "{Search for Dark-Matter-Induced Oscillations of Fundamental Constants Using Molecular Spectroscopy}",
    eprint = "2111.06883",
    archivePrefix = "arXiv",
    primaryClass = "hep-ph",
    doi = "10.1103/PhysRevLett.129.031302",
    journal = "Phys. Rev. Lett.",
    volume = "129",
    number = "3",
    pages = "031302",
    year = "2022"
}

@article{Banks:2024sli,
    author = "Banks, Hannah and Fuchs, Elina and McCullough, Matthew",
    title = "{A Nuclear Interferometer for Ultra-Light Dark Matter Detection}",
    eprint = "2407.11112",
    archivePrefix = "arXiv",
    primaryClass = "hep-ph",
    reportNumber = "CERN-TH-2024-104",
    doi = "10.1103/ntb7-6w5g",
    month = "7",
    year = "2024"
}

@article{Planck:2018vyg,
    author = "Aghanim, N. and others",
    collaboration = "Planck",
    title = "{Planck 2018 results. VI. Cosmological parameters}",
    eprint = "1807.06209",
    archivePrefix = "arXiv",
    primaryClass = "astro-ph.CO",
    doi = "10.1051/0004-6361/201833910",
    journal = "Astron. Astrophys.",
    volume = "641",
    pages = "A6",
    year = "2020",
    note = "[Erratum: Astron.Astrophys. 652, C4 (2021)]"
}

@article{Hu:2000ke,
    author = "Hu, Wayne and Barkana, Rennan and Gruzinov, Andrei",
    title = "{Cold and fuzzy dark matter}",
    eprint = "astro-ph/0003365",
    archivePrefix = "arXiv",
    doi = "10.1103/PhysRevLett.85.1158",
    journal = "Phys. Rev. Lett.",
    volume = "85",
    pages = "1158--1161",
    year = "2000"
}

@article{Antypas:2022asj,
    author = "Antypas, D. and others",
    title = "{New Horizons: Scalar and Vector Ultralight Dark Matter}",
    eprint = "2203.14915",
    archivePrefix = "arXiv",
    primaryClass = "hep-ex",
    reportNumber = "FERMILAB-PUB-22-262-AD-PPD-T",
    month = "3",
    year = "2022",
    journal=""
}

@article{Hui:2021tkt,
    author = "Hui, Lam",
    title = "{Wave Dark Matter}",
    eprint = "2101.11735",
    archivePrefix = "arXiv",
    primaryClass = "astro-ph.CO",
    doi = "10.1146/annurev-astro-120920-010024",
    journal = "Ann. Rev. Astron. Astrophys.",
    volume = "59",
    pages = "247--289",
    year = "2021"
}

@article{Khoury:2025txd,
    author = "Khoury, Justin and Lin, Meng-Xiang and Trodden, Mark",
    title = "{Apparent w{\ensuremath{<}}-1 and a Lower S8 from Dark Axion and Dark Baryons Interactions}",
    eprint = "2503.16415",
    archivePrefix = "arXiv",
    primaryClass = "astro-ph.CO",
    doi = "10.1103/w4qb-plk8",
    journal = "Phys. Rev. Lett.",
    volume = "135",
    number = "18",
    pages = "181001",
    year = "2025"
}

@article{Erickcek:2013oma,
    author = "Erickcek, Adrienne L. and Barnaby, Neil and Burrage, Clare and Huang, Zhiqi",
    title = "{Catastrophic Consequences of Kicking the Chameleon}",
    eprint = "1304.0009",
    archivePrefix = "arXiv",
    primaryClass = "astro-ph.CO",
    doi = "10.1103/PhysRevLett.110.171101",
    journal = "Phys. Rev. Lett.",
    volume = "110",
    pages = "171101",
    year = "2013"
}

@article{Hook:2017psm,
    author = "Hook, Anson and Huang, Junwu",
    title = "{Probing axions with neutron star inspirals and other stellar processes}",
    eprint = "1708.08464",
    archivePrefix = "arXiv",
    primaryClass = "hep-ph",
    doi = "10.1007/JHEP06(2018)036",
    journal = "JHEP",
    volume = "06",
    pages = "036",
    year = "2018"
}

@article{Alarcon:2011zs,
    author = "Alarcon, J. M. and Martin Camalich, J. and Oller, J. A.",
    title = "{The chiral representation of the $\pi N$ scattering amplitude and the pion-nucleon sigma term}",
    eprint = "1110.3797",
    archivePrefix = "arXiv",
    primaryClass = "hep-ph",
    doi = "10.1103/PhysRevD.85.051503",
    journal = "Phys. Rev. D",
    volume = "85",
    pages = "051503",
    year = "2012"
}

@article{Saikawa:2018rcs,
    author = "Saikawa, Ken'ichi and Shirai, Satoshi",
    title = "{Primordial gravitational waves, precisely: The role of thermodynamics in the Standard Model}",
    eprint = "1803.01038",
    archivePrefix = "arXiv",
    primaryClass = "hep-ph",
    reportNumber = "IPMU18-0037, MPP-2018-19",
    doi = "10.1088/1475-7516/2018/05/035",
    journal = "JCAP",
    volume = "05",
    pages = "035",
    year = "2018"
}

@article{Sibiryakov:2020eir,
    author = "Sibiryakov, Sergey and S{\o}rensen, Philip and Yu, Tien-Tien",
    title = "{BBN constraints on universally-coupled ultralight scalar dark matter}",
    eprint = "2006.04820",
    archivePrefix = "arXiv",
    primaryClass = "hep-ph",
    reportNumber = "DESY-19-234, CERN-TH-2020-091, INR-TH-2020-001",
    doi = "10.1007/JHEP12(2020)075",
    journal = "JHEP",
    volume = "12",
    pages = "075",
    year = "2020"
}

@article{Hook:2018jle,
    author = "Hook, Anson",
    title = "{Solving the Hierarchy Problem Discretely}",
    eprint = "1802.10093",
    archivePrefix = "arXiv",
    primaryClass = "hep-ph",
    doi = "10.1103/PhysRevLett.120.261802",
    journal = "Phys. Rev. Lett.",
    volume = "120",
    number = "26",
    pages = "261802",
    year = "2018"
}

@article{Ferreira:2020fam,
    author = "Ferreira, Elisa G. M.",
    title = "{Ultra-light dark matter}",
    eprint = "2005.03254",
    archivePrefix = "arXiv",
    primaryClass = "astro-ph.CO",
    doi = "10.1007/s00159-021-00135-6",
    journal = "Astron. Astrophys. Rev.",
    volume = "29",
    number = "1",
    pages = "7",
    year = "2021"
}

@article{OHare:2024nmr,
    author = "O'Hare, Ciaran A. J.",
    title = "{Cosmology of axion dark matter}",
    eprint = "2403.17697",
    archivePrefix = "arXiv",
    primaryClass = "hep-ph",
    doi = "10.22323/1.454.0040",
    journal = "PoS",
    volume = "COSMICWISPers",
    pages = "040",
    year = "2024"
}

@article{Hees:2018fpg,
    author = "Hees, Aur{\'e}lien and Minazzoli, Olivier and Savalle, Etienne and Stadnik, Yevgeny V. and Wolf, Peter",
    title = "{Violation of the equivalence principle from light scalar dark matter}",
    eprint = "1807.04512",
    archivePrefix = "arXiv",
    primaryClass = "gr-qc",
    doi = "10.1103/PhysRevD.98.064051",
    journal = "Phys. Rev. D",
    volume = "98",
    number = "6",
    pages = "064051",
    year = "2018"
}

@article{Damour:2010rp,
    author = "Damour, Thibault and Donoghue, John F.",
    title = "{Equivalence Principle Violations and Couplings of a Light Dilaton}",
    eprint = "1007.2792",
    archivePrefix = "arXiv",
    primaryClass = "gr-qc",
    doi = "10.1103/PhysRevD.82.084033",
    journal = "Phys. Rev. D",
    volume = "82",
    pages = "084033",
    year = "2010"
}

@article{Damour:2010rm,
    author = "Damour, Thibault and Donoghue, John F.",
    title = "{Phenomenology of the Equivalence Principle with Light Scalars}",
    eprint = "1007.2790",
    archivePrefix = "arXiv",
    primaryClass = "gr-qc",
    doi = "10.1088/0264-9381/27/20/202001",
    journal = "Class. Quant. Grav.",
    volume = "27",
    pages = "202001",
    year = "2010"
}

@article{Lee:2020zjt,
    author = "Lee, J. G. and Adelberger, E. G. and Cook, T. S. and Fleischer, S. M. and Heckel, B. R.",
    title = "{New Test of the Gravitational $1/r^2$ Law at Separations down to 52 $\mu$m}",
    eprint = "2002.11761",
    archivePrefix = "arXiv",
    primaryClass = "hep-ex",
    doi = "10.1103/PhysRevLett.124.101101",
    journal = "Phys. Rev. Lett.",
    volume = "124",
    number = "10",
    pages = "101101",
    year = "2020"
}

@article{Bauer:2023czj,
    author = "Bauer, Martin and Rostagni, Guillaume",
    title = "{Fifth Forces from QCD Axions Scale Differently}",
    eprint = "2307.09516",
    archivePrefix = "arXiv",
    primaryClass = "hep-ph",
    reportNumber = "IPPP/23/35",
    doi = "10.1103/PhysRevLett.132.101802",
    journal = "Phys. Rev. Lett.",
    volume = "132",
    number = "10",
    pages = "101802",
    year = "2024"
}

@article{Grossman:2025cov,
    author = "Grossman, Yuval and Yu, Bingrong and Zhou, Siyu",
    title = "{Axion forces in axion backgrounds}",
    eprint = "2504.00104",
    archivePrefix = "arXiv",
    primaryClass = "hep-ph",
    doi = "10.1007/JHEP01(2026)145",
    journal = "JHEP",
    volume = "01",
    pages = "145",
    year = "2026"
}

@article{OConnell:2006rsp,
    author = "O'Connell, Donal and Ramsey-Musolf, Michael J. and Wise, Mark B.",
    title = "{Minimal Extension of the Standard Model Scalar Sector}",
    eprint = "hep-ph/0611014",
    archivePrefix = "arXiv",
    reportNumber = "CALT-68-2614",
    doi = "10.1103/PhysRevD.75.037701",
    journal = "Phys. Rev. D",
    volume = "75",
    pages = "037701",
    year = "2007"
}

@article{Patt:2006fw,
    author = "Patt, Brian and Wilczek, Frank",
    title = "{Higgs-field portal into hidden sectors}",
    eprint = "hep-ph/0605188",
    archivePrefix = "arXiv",
    reportNumber = "MIT-CTP-3745",
    month = "5",
    year = "2006"
}

@article{Delaunay:2025pho,
    author = "Delaunay, C{\'e}dric and Geller, Michael and Heller-Algazi, Zamir and Perez, Gilad and Springmann, Konstantin",
    title = "{Natural Ultralight Dark Matter: The Quadratic Twin}",
    eprint = "2507.12514",
    archivePrefix = "arXiv",
    primaryClass = "hep-ph",
    month = "7",
    year = "2025"
}

@article{Burrage:2018dvt,
    author = "Burrage, Clare and Copeland, Edmund J. and Millington, Peter and Spannowsky, Michael",
    title = "{Fifth forces, Higgs portals and broken scale invariance}",
    eprint = "1804.07180",
    archivePrefix = "arXiv",
    primaryClass = "hep-th",
    reportNumber = "IPPP/18/23, IPPP-18-23",
    doi = "10.1088/1475-7516/2018/11/036",
    journal = "JCAP",
    volume = "11",
    pages = "036",
    year = "2018"
}

@article{Erickcek:2013dea,
    author = "Erickcek, Adrienne L. and Barnaby, Neil and Burrage, Clare and Huang, Zhiqi",
    title = "{Chameleons in the Early Universe: Kicks, Rebounds, and Particle Production}",
    eprint = "1310.5149",
    archivePrefix = "arXiv",
    primaryClass = "astro-ph.CO",
    doi = "10.1103/PhysRevD.89.084074",
    journal = "Phys. Rev. D",
    volume = "89",
    number = "8",
    pages = "084074",
    year = "2014"
}

@article{Padilla:2015wlv,
    author = "Padilla, Antonio and Platts, Emma and Stefanyszyn, David and Walters, Anthony and Weltman, Amanda and Wilson, Toby",
    title = "{How to Avoid a Swift Kick in the Chameleons}",
    eprint = "1511.05761",
    archivePrefix = "arXiv",
    primaryClass = "hep-th",
    doi = "10.1088/1475-7516/2016/03/058",
    journal = "JCAP",
    volume = "03",
    pages = "058",
    year = "2016"
}

@article{Brax:2004qh,
    author = "Brax, Philippe and van de Bruck, Carsten and Davis, Anne-Christine and Khoury, Justin and Weltman, Amanda",
    title = "{Detecting dark energy in orbit: The cosmological chameleon}",
    eprint = "astro-ph/0408415",
    archivePrefix = "arXiv",
    doi = "10.1103/PhysRevD.70.123518",
    journal = "Phys. Rev. D",
    volume = "70",
    pages = "123518",
    year = "2004"
}

@article{Damour:1994ya,
    author = "Damour, T. and Polyakov, Alexander M.",
    title = "{String theory and gravity}",
    eprint = "gr-qc/9411069",
    archivePrefix = "arXiv",
    reportNumber = "IHES-P-94-1",
    doi = "10.1007/BF02106709",
    journal = "Gen. Rel. Grav.",
    volume = "26",
    pages = "1171--1176",
    year = "1994"
}

@article{Damour:1993id,
    author = "Damour, T. and Nordtvedt, K.",
    title = "{Tensor - scalar cosmological models and their relaxation toward general relativity}",
    reportNumber = "IHES-P-93-16",
    doi = "10.1103/PhysRevD.48.3436",
    journal = "Phys. Rev. D",
    volume = "48",
    pages = "3436--3450",
    year = "1993"
}

@article{Damour:1994zq,
    author = "Damour, T. and Polyakov, Alexander M.",
    title = "{The String dilaton and a least coupling principle}",
    eprint = "hep-th/9401069",
    archivePrefix = "arXiv",
    doi = "10.1016/0550-3213(94)90143-0",
    journal = "Nucl. Phys. B",
    volume = "423",
    pages = "532--558",
    year = "1994"
}

@article{Dondi:2019olm,
    author = "Dondi, Nicola Andrea and Sannino, Francesco and Smirnov, Juri",
    title = "{Thermal history of composite dark matter}",
    eprint = "1905.08810",
    archivePrefix = "arXiv",
    primaryClass = "hep-ph",
    doi = "10.1103/PhysRevD.101.103010",
    journal = "Phys. Rev. D",
    volume = "101",
    number = "10",
    pages = "103010",
    year = "2020"
}

@article{Garani:2021zrr,
    author = "Garani, Raghuveer and Redi, Michele and Tesi, Andrea",
    title = "{Dark QCD matters}",
    eprint = "2105.03429",
    archivePrefix = "arXiv",
    primaryClass = "hep-ph",
    doi = "10.1007/JHEP12(2021)139",
    journal = "JHEP",
    volume = "12",
    pages = "139",
    year = "2021"
}

@article{Tsai:2020vpi,
    author = "Tsai, Yu-Dai and McGehee, Robert and Murayama, Hitoshi",
    title = "{Resonant Self-Interacting Dark Matter from Dark QCD}",
    eprint = "2008.08608",
    archivePrefix = "arXiv",
    primaryClass = "hep-ph",
    reportNumber = "FERMILAB-PUB-20-365-AE-T",
    doi = "10.1103/PhysRevLett.128.172001",
    journal = "Phys. Rev. Lett.",
    volume = "128",
    number = "17",
    pages = "172001",
    year = "2022"
}

@article{Morrison:2020yeg,
    author = "Morrison, Logan and Profumo, Stefano and Robinson, Dean J.",
    title = "{Large $N$-ightmare Dark Matter}",
    eprint = "2010.03586",
    archivePrefix = "arXiv",
    primaryClass = "hep-ph",
    doi = "10.1088/1475-7516/2021/05/058",
    journal = "JCAP",
    volume = "05",
    pages = "058",
    year = "2021"
}

@article{GrillidiCortona:2015jxo,
    author = "Grilli di Cortona, Giovanni and Hardy, Edward and Pardo Vega, Javier and Villadoro, Giovanni",
    title = "{The QCD axion, precisely}",
    eprint = "1511.02867",
    archivePrefix = "arXiv",
    primaryClass = "hep-ph",
    doi = "10.1007/JHEP01(2016)034",
    journal = "JHEP",
    volume = "01",
    pages = "034",
    year = "2016"
}

@article{DiVecchia:1980yfw,
    author = "Di Vecchia, P. and Veneziano, G.",
    title = "{Chiral Dynamics in the Large n Limit}",
    reportNumber = "CERN-TH-2814",
    doi = "10.1016/0550-3213(80)90370-3",
    journal = "Nucl. Phys. B",
    volume = "171",
    pages = "253--272",
    year = "1980"
}

@article{SevillanoMunoz:2024ayh,
    author = "Sevillano Mu{\~n}oz, Sergio",
    title = "{A particle's perspective on screening mechanisms}",
    eprint = "2407.08779",
    archivePrefix = "arXiv",
    primaryClass = "hep-ph",
    reportNumber = "IPPP/24/45",
    doi = "10.1088/1475-7516/2024/12/052",
    journal = "JCAP",
    volume = "12",
    pages = "052",
    year = "2024"
}

@article{Batell:2021ofv,
    author = "Batell, Brian and Ghalsasi, Akshay",
    title = "{Thermal misalignment of scalar dark matter}",
    eprint = "2109.04476",
    archivePrefix = "arXiv",
    primaryClass = "hep-ph",
    reportNumber = "PITT-PACC-2119",
    doi = "10.1103/PhysRevD.107.L091701",
    journal = "Phys. Rev. D",
    volume = "107",
    number = "9",
    pages = "L091701",
    year = "2023"
}

@article{Batell:2022qvr,
    author = "Batell, Brian and Ghalsasi, Akshay and Rai, Mudit",
    title = "{Dynamics of dark matter misalignment through the Higgs portal}",
    eprint = "2211.09132",
    archivePrefix = "arXiv",
    primaryClass = "hep-ph",
    reportNumber = "PITT-PACC-2213",
    doi = "10.1007/JHEP01(2024)038",
    journal = "JHEP",
    volume = "01",
    pages = "038",
    year = "2024"
}

@article{Cyncynates:2024ufu,
    author = "Cyncynates, David and Simon, Olivier",
    title = "{Scalar Relics from the Hot Big Bang}",
    eprint = "2410.22409",
    archivePrefix = "arXiv",
    primaryClass = "hep-ph",
    doi = "10.1103/wjt8-9dh9",
    journal = "Phys. Rev. Lett.",
    volume = "135",
    number = "10",
    pages = "101003",
    year = "2025"
}

@article{Cyncynates:2024bxw,
    author = "Cyncynates, David and Simon, Olivier",
    title = "{Minimal targets for dilaton direct detection}",
    eprint = "2408.16816",
    archivePrefix = "arXiv",
    primaryClass = "hep-ph",
    doi = "10.1103/8mc9-6cmr",
    journal = "Phys. Rev. D",
    volume = "112",
    number = "5",
    pages = "055002",
    year = "2025"
}

@article{GuptaChoudhury:2025uje,
    author = "Gupta Choudhury, Shibendu and Dutta, Koushik and Ghosh, Deep",
    title = "{Scalaron dark matter dynamics: effects of Higgs non-minimal coupling to gravity}",
    eprint = "2509.21148",
    archivePrefix = "arXiv",
    primaryClass = "hep-ph",
    month = "9",
    year = "2025"
}

@article{Ferrante:2025avp,
    author = "Ferrante, Steven and Perelstein, Maxim and Yu, Bingrong",
    title = "{WIMP Meets ALP: Coherent Freeze-Out of Dark Matter}",
    eprint = "2511.16731",
    archivePrefix = "arXiv",
    primaryClass = "hep-ph",
    month = "11",
    year = "2025"
}

@misc{SevillanoMunoz:Code,
  author       = {Burrage, Clare and Sevillano Mu{\~n}oz, Sergio},
  title        = {{Misalignment Kicks}},
  year         = {2026},
  howpublished = {\href{https://github.com/SergioSevi/Misalignment_kicks}{GitHub repository}}
}

@article{Preskill:1982cy,
    author = "Preskill, John and Wise, Mark B. and Wilczek, Frank",
    editor = "Srednicki, M. A.",
    title = "{Cosmology of the Invisible Axion}",
    reportNumber = "HUTP-82-A048, NSF-ITP-82-103",
    doi = "10.1016/0370-2693(83)90637-8",
    journal = "Phys. Lett. B",
    volume = "120",
    pages = "127--132",
    year = "1983"
}

@article{Abbott:1982af,
    author = "Abbott, L. F. and Sikivie, P.",
    editor = "Srednicki, M. A.",
    title = "{A Cosmological Bound on the Invisible Axion}",
    reportNumber = "PRINT-82-0695 (BRANDEIS)",
    doi = "10.1016/0370-2693(83)90638-X",
    journal = "Phys. Lett. B",
    volume = "120",
    pages = "133--136",
    year = "1983"
}

@article{Dine:1982ah,
    author = "Dine, Michael and Fischler, Willy",
    editor = "Srednicki, M. A.",
    title = "{The Not So Harmless Axion}",
    reportNumber = "UPR-0201T",
    doi = "10.1016/0370-2693(83)90639-1",
    journal = "Phys. Lett. B",
    volume = "120",
    pages = "137--141",
    year = "1983"
}

\end{document}